\documentclass[epj,final]{svjour}
\usepackage{graphics}
\usepackage{graphicx}

\begin{document}

\title{Dipolar interaction and incoherent quantum tunneling: a Monte Carlo
  study of magnetic relaxation }

\author{A. Cuccoli\inst{1} \and A. Fort\inst{2} \and A. Rettori \inst{1} \and
E. Adam\inst{3} \and J. Villain\inst{3,}\inst{4}}

\institute{
  Dipartimento di Fisica dell'Universit{\`a} di Firenze
    and Istituto Nazionale di Fisica della Materia, \\
    Largo E. Fermi 2, I-50125 Firenze, Italy \and 
  Istituto di Elettronica Quantistica del CNR - Via
    Panciatichi 56/30, I-50127 Firenze, Italy \and
  D{\'e}partement de Recherche Fondamentale sur la Mati{\'e}re 
    Condens{\'e}e, CEA Grenoble, \\
    17 Avenue des Martyrs, F-38054 Grenoble Cedex 9, France \and
  CRTBT, CNRS, B.P. 166, F-38042 Grenoble Cedex 9
} 

\date{\today}

\abstract{
  We study the magnetic relaxation of a system of localized spins
  interacting through weak dipole interactions, at a temperature large
  with respect to the ordering temperature but low with respect to the
  crystal field level splitting.  The relaxation results from quantum
  spin tunneling but is only allowed on sites where the dipole field
  is very small. At low times, the magnetization decrease is
  proportional to $\sqrt{t}$ as predicted by Prokofiev and Stamp, and
  at long times the relaxation can be described as an extension of a
  relaxed zone. The results can be directly compared with very recent
  experimental data on Fe$_8$ molecular clusters.
} 

\maketitle

\section{A novel relaxation mechanism}
\label{ch1}

The relaxation of a quantity $m(t)$ toward its equilibrium value $m(\infty)$
is usually described by a standard model taken from a surprisingly limited set,
in contrast with the extreme diversity of the physical problems to which each 
of these models apply. The simplest model is the linear equation

\begin{equation}
\frac{dm(t)}{dt} = -\alpha [m(t)-m(\infty)]
\label{1.1}
\end{equation}
where $\alpha$ is a constant. 

The compound [(tacn)$_8$Fe$_8$O$_8$(OH)$_8$]$^{8+}$ (where tacn is
1-4-7-triazacyclononane) pertains, at low temperature T $<$ 1K, to a new class
which has not been studied until very recently~\cite{prok,prok'}.  This
material~\cite{Sangrego}, hereafter called `Fe$_8$', is a paramagnet
and the quantity $m(t)$ of interest is the magnetization per spin,
more precisely its component along a well defined axis $z$ which is an
easy magnetization axis. According to current
knowledge~\cite{Sangrego},
this material is made of molecular groups, each of which contains 8
ions Fe$^{+++}$ and has at low temperature a spin $s=10$ which results
from a strong exchange coupling between the 8 Fe$^{+++}$ ions. At low
temperature each molecular spin $\vec S_i$ is oriented along $z$,
with the value
\begin{equation}
 S_i^z=S_i= \pm s~.
\label{1.1'}
\end{equation}

The weak dipole 
interaction between different molecular groups is not sufficient to produce
a magnetic order at any accessible temperature and,
at thermal equilibrium, the +s and -s spins are randomly 
distributed with an average value $m(\infty)$ which depends on
the external field.  
In a typical relaxation experiment at very low temperature, all 
spins are initially in the state 

\begin{equation}
 S_i^z(0) = -s
\label{1.1''}
\end{equation}
so that $m(0)=-s$.

Above 1 K, the relaxation is well described by (\ref{1.1})
and the relaxation is exponential~\cite{Sangrego},

\begin{equation}
m(t) - m(\infty) = [m(0) - m(\infty)]\exp(-t/\tau)~.
\label{1.2}
\end{equation}

However, at low temperature, the relaxation is not exponential. At long
times $t$, it is pretty
well described by a stretched exponential~\cite{Sangrego} 

\begin{equation}
m(t) - m(\infty) = [m(0) - m(\infty)]\exp[-(t/\tau)^{ \beta_1}]
\label{1.3}
\end{equation}
where the exponent $ \beta_1$ depends on $T$ and takes the value 1
above 1 K.  Below 0.4 K, $ \beta_1$ is nearly
constant~\cite{Sangrego} and equal to

\begin{equation}
\beta_1 \approx 0.4~.
\label{1.4}
\end{equation}

In the present work, kinetic Monte-Carlo simulations of the magnetic  
relaxation of Fe$_8$ are reported; they are made on the basis of a
model described in the next section.

\section{The incoherent tunneling model}
\label{ch2}

At the low temperatures of interest, thermal activation is impossible and  
the relaxation takes place by tunneling. In weak external field,
tunneling takes place between the two states (\ref{1.1'}), and  
is only possible for a  spin $\vec S_i$ 
if the $z$-component $H_i$ of the local magnetic field is very close to 0, say 
\begin{equation}
-H_1 < H_i < H_1
\label{2.1}
\end{equation}
where $H_1$ is a constant which will be precised below.  The local
field is the sum of the external field $H_{ext}$ and an internal
field. In most of materials, the internal field is partly due to
nuclear spins (hyperfine interactions).  When the magnetic particles
are Fe, the hyperfine contact interaction is nearly absent because the
most common Fe isotope has no spin, the magnetic isotope has a weak
concentration (2\%) and moreover a weak spin (1/2). In the present
work, hyperfine interactions will not be explicitly taken into account
in the evaluation of the internal field. They are but implicitly
included in the theory since they are probably~\cite{prok,prok'}
responsible for the resonance width $H_1$.  This width has been
measured by Wernsdorfer et al.~\cite{wern} in Fe$_8$ and is of order
10 Oersteds.

Thus, the internal field only 
depends on the molecular spins through the formula 

\begin{equation}
H^{(i)}_{dip}= \sum _j g_{ij}S_j~.
\label{2.1'}
\end{equation}
This field will hereafter be called `dipole field' although it may
also contain an exchange component. At long distance this exchange part 
vanishes and the 
coefficients $g_{ij}=g(\vec r_i - \vec r_j)$ behave  according 
to the formula
\begin{equation}
g(\vec r)= -\frac{K}{r^{3}}\left( 1 - 3z^2/r^2 \right)
\label{2.2}
\end{equation}
where $z$ is the component of $\vec r$ on the $z$ axis. 

In a finite system, tunneling is an undamped oscillation between two
states and does not really result in relaxation. For instance, if the
spin $\vec S_i$ were isolated, it would oscillate in zero field
between the two states (\ref{1.1'}).  This is not true for the large
systems which are studied in experimental physics, and it is
reasonable to make the following assumption. 

\vskip5mm
{\it Basic assumption.}
\vskip3mm

Any spin $i$ subject to
a  local field $H_i$  
has a  probability $\eta(H_i)$ per unit time of transition
between the two states (\ref{1.1'}). The real, nonnegative function 
$\eta(H_i)$ is negligible if
$H_i$ does not satisfy (\ref{2.1}).

\vskip5mm

The choice of the function $\eta(H)$ is presumably not essential provided 
the above properties are satisfied. A possible choice is

\begin{eqnarray}
\left\{
\begin{array}{lll}
\displaystyle{
\eta(H)= \eta(0) \left[ 1 - \frac{H^2}{H_1^2}  \right]   
 \;\;\;\;(-H_1<H<H_1)
}&\\
\displaystyle{ }&\\
\displaystyle{
\eta(H)= 0\;\;\;\;(|H|>H_1)
  }~.&  \\
\end{array}
\right.
\label{2.3}
\end{eqnarray}

The above  assumption defines the `incoherent tunneling model'
which is studied in the next sections.
Since the spin-flip transitions
modify the fields, the relaxation defined by the basic assumptions  and 
by formula (\ref{2.1'}) is difficult. It
has been investigated by kinetic Monte Carlo simulations by  
Prokofiev \& Stamp~\cite{prok} and by Ohm \& Paulsen~\cite{ohm}. 
More detailed results are presented in  section \ref{ch5}.
They include a detailed analysis of the effect of the
sample shape and crystal structure. 

Prokofiev and Stamp have 
given an analytic description of the short time behavior.
A critical summary of this theory is given in the next section.

In higher external field, the basic assumption must be modified since 
tunneling becomes possible from state $S^i_z=-s$ to an eigenstate
(or rather, nearly eigenstate) $S^i_z=m$, with $0<m<s$, if the external field 
is such that two eigenvectors of the spin Hamiltonian have nearly the same 
energy (resonance condition)~\cite{HB}. Then, the spin 
emits phonons and goes to state $S^i_z=s$. It would be easy to include 
this possibility in the following calculations, but for the
sake of simplicity it will be ignored. This is correct if the external field 
is not too large.

\section{The Prokofiev-Stamp approximation for a sphere}
\label{ch3}
                                                            
Since the dipole interactions (\ref{2.1'}) are essential in the model, 
the shape of the sample should be important if the model is correct.
The simplest case is that of a spherical sample. Then (\ref{1.1''})
implies, at $t=0$, that the local field $H_i$ has the same 
value $H(0)$ for all spins $i$,
{\it except} near the surface. 

It will be assumed in this section that the sample is spherical and
that the external field is such that $H(0)=0$. These conditions allow
the rapid reversal of a certain amount of spins. Because of this
process, the local field is no longer uniform, so that the reversal of
most of the other spins is hindered. The distribution $P(t;H)$ of the
local fields at time $t$ will be assumed to be continuous, and the
proportion of spins which can reverse in the time unit is of order
$\eta(0) H_1 P(t;0)$; this value is exact if $\eta(H)=$const. in a
range of width $H_1$ and zero outside.
If one assumes that the local field $H_i$ and the spin $S_i$ are
independent at each time, it follows from Eq. (10) 
\begin{equation}
\frac{d}{dt}m(t)=-\frac{4}{3}\eta(0) H_1  P(t;0)m(t)~.
\label{3.1}
\end{equation}

The function  $P(t;H)$ is characterized by its width $\Gamma(t)$,
so that (\ref{3.1}) can be replaced by 

\begin{equation}
\frac{d}{dt}m(t)=-\frac{4}{3}\eta(0) H_1 m(t)/ \Gamma(t)~. 
\label{3.2}
\end{equation}

In order to obtain a closed evolution equation for $m(t)$, one should
relate $\Gamma(t)$ to the magnetization $m(t)$.  When $m$ has its
saturation value $-s$, as is the case at $t=0$, the width $\Gamma$
vanishes for a spherical sample.  According to Prokofiev and Stamp,
$\Gamma(t)=\Gamma(m(t))$ with

\begin{equation}
 \Gamma(m) = (1-|m|/s) H_d 
\label{3.3}
\end{equation}
where the constant $H_d$ has the order of magnitude of the 
maximum dipole field.

Relation (\ref{3.3}) can be justified as follows. At
short times, the local field is the
sum of the initial field, which has been assumed to be 0, and the 
dipole field produced by those spins which are already reversed. 
Let $\ell$ be the average distance between those spins. The typical 
value of the resulting
dipole field, as given by (\ref{2.2}), is of order 
$K/\ell^3\approx H_d a^3/\ell^3$,
where $a$ is the distance between spins.
The width $\Gamma(m)$ should have the same order of magnitude, 
$\Gamma(m)\approx H_d a^3/\ell^3$.
Now, $a^3/\ell^3$ is the proportion of reversed spins, i.e. $(1-|m|/s)/2$.
Relation (\ref{3.3}) follows. 

Insertion of (\ref{3.3}) into (\ref{3.2}) yields

\begin{equation}
[1-|m(t)|/s]\frac{d}{dt}[1-|m(t)|/s]\approx \eta(0) H_1 |m(t)|/ (sH_d )
\label{3.4}
\end{equation}
the solution of which is

\begin{equation}
\eta(0) H_1 t/ H_d \approx |m(t)|/s - 1 -\ln [|m(t)|/s]~.
\label{3.4'}
\end{equation}

For short times, (\ref{3.4'}) reduces to the result of Prokofiev and
Stamp
 \cite{prok}
\begin{equation}
1-|m(t)|/s \approx \sqrt{t/ \tau_a }
\label{3.5}
\end{equation}
while, for long times, (\ref{3.4'}) reduces  to
                        
\begin{equation}
|m(t)|/s \approx \exp \left[ -\frac{ t}{ \tau_a} -1 \right]
\label{3.6}
\end{equation}          
where
\begin{equation}
 \tau_a = \frac{H_d}{\eta(0) H_1 }~.
\label{3.6'}
\end{equation}          

Formula (\ref{3.5}) will be seen to describe correctly the short time behaviour,
while the long time behaviour will be seen to be described by (\ref{1.3}). 
Formulae (\ref{3.4'})
and (\ref{3.6}) will be seen not to be satisfactory at long times. This 
remark shows that the approximations made in the derivation of  (\ref{3.5})
are only valid at short times.

\section{Non-resonant fields and non-spherical samples}
\label{ch3'}

The analytic formulae (\ref{3.4'}) to (\ref{3.6}) have been obtained for a
spherical sample when the external field is such that, at $t=0$, the `resonance
condition' $H_i=0$ is satisfied for all spins $i$.

\begin{figure}[tb]
\begin{center}
\includegraphics[bbllx=40pt,bblly=40pt,bburx=520pt,bbury=440pt,
width=7 truecm ,angle=0]{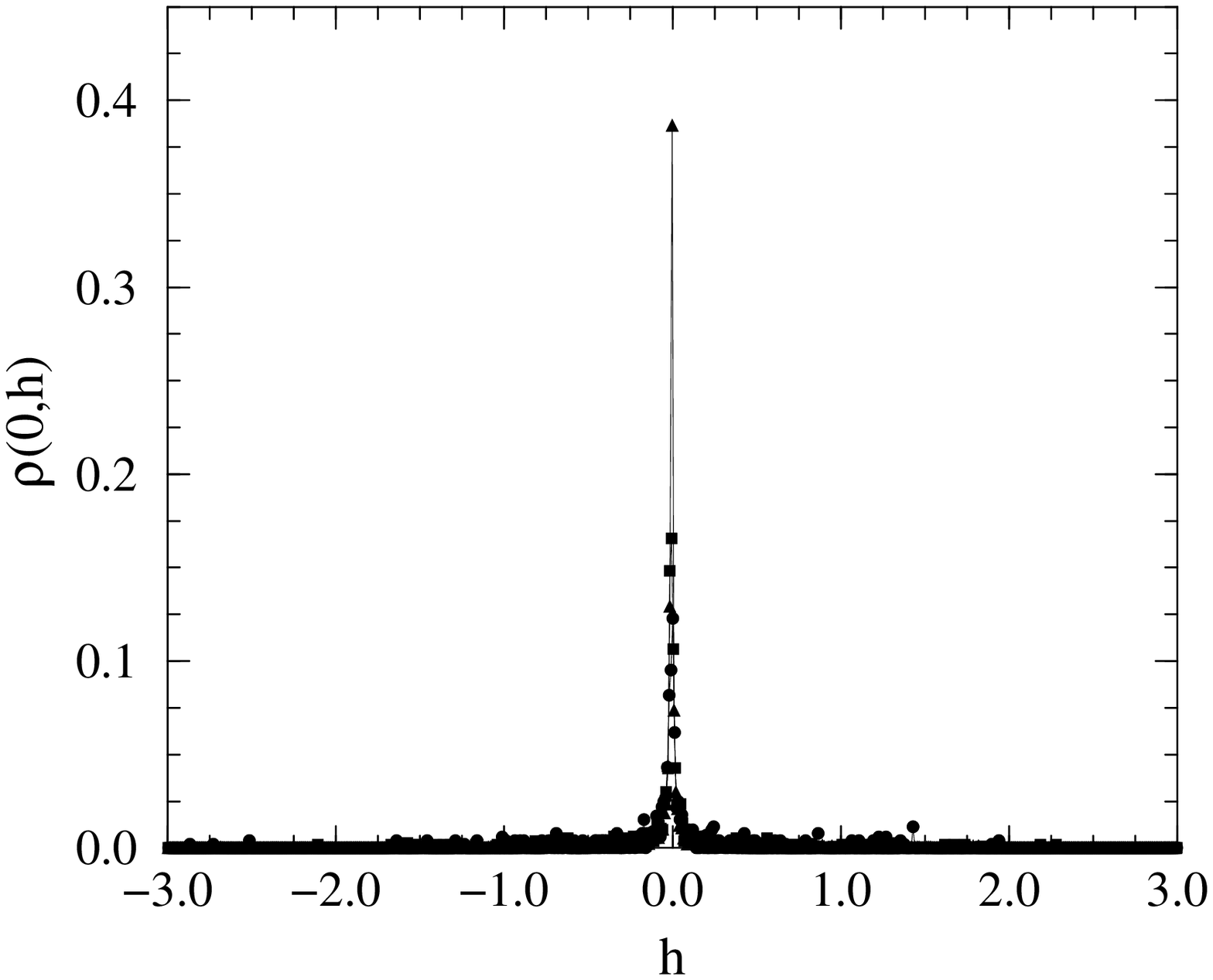}
\end{center}
\begin{center}
\includegraphics[bbllx=0pt,bblly=0pt,bburx=440pt,bbury=465pt,
width=6 truecm,angle=0]{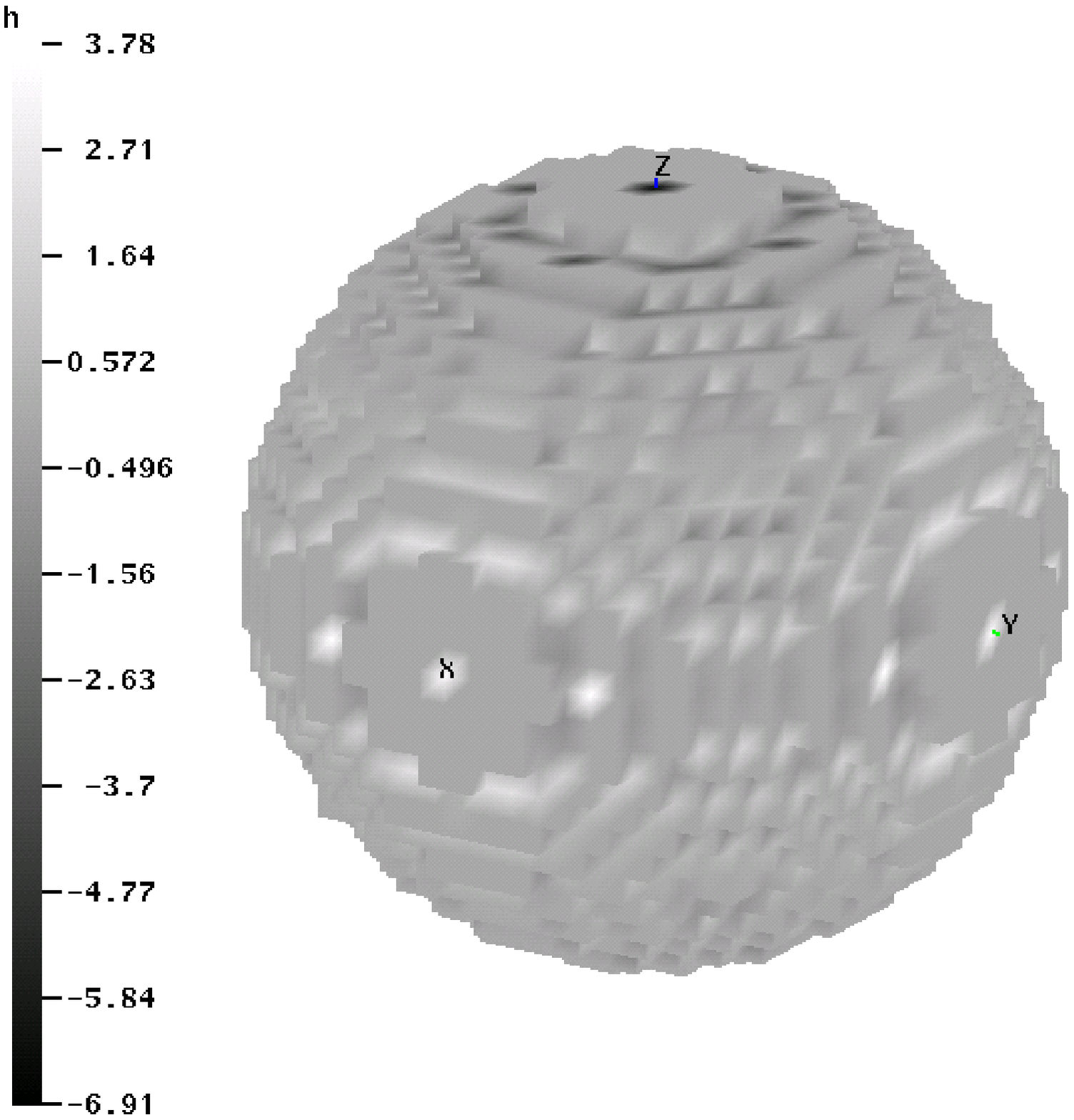} 
\end{center}
\nobreak\hspace{-2 truecm}
\begin{center}
\includegraphics[bbllx=45pt,bblly=85pt,bburx=565pt,bbury=380pt,
width=7 truecm ,angle=0]{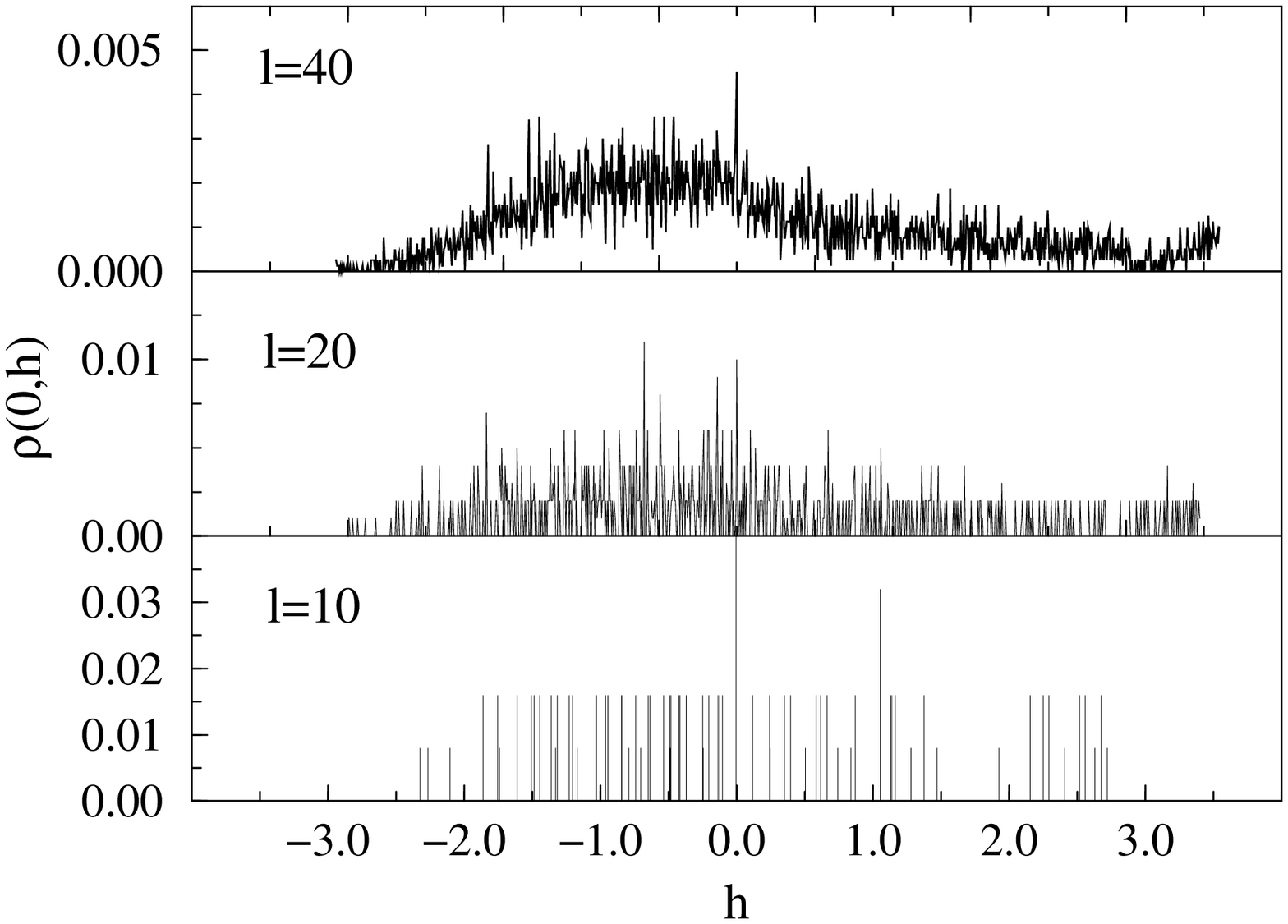}
\end{center}
\caption{Distribution of local fields
at the sites of a simple cubic
lattice occupied by spins parallel to the $z$ axis. a)
Sphere of 9134 spins. The wings are due to the surface,
as shown in picture b). In such picture the clear regions are those 
where the local field is 
larger than its average value, the dark regions are those where 
it is lower. 
c) Cube of different sizes: $10 \times 10 \times 10 $, 
$20 \times 20 \times 20 $, and $40 \times 40 \times 40 $ spins,
starting from below. The field is measured in units $g\mu_BS/a^3$
($\simeq$ 80 Oersted, setting $S=10$ and $a\simeq 13~\mbox{\AA}$, 
average distance among Fe$_8$ clusters in the actual compound). Note
as the size of the sample affects the discretization of the field values.}
\label{t=0}
\end{figure}

If in (\ref{2.1'}) the sum is replaced by an integral, the local field
is uniform in a spherical sample, and vanishes exactly at saturation
(i.e. if $S_i=-s$) if the external field is absent. However, 
the replacement by an integral is not
exact, and the local field is not uniform at saturation, although it
is very sharply peaked (Fig. \ref{t=0}a). As a matter of fact, it is
impossible to cut a perfect sphere in a crystalline material, and the
edges can only be approximately spherical~\cite{OhmPhD}.

Relaxation can thus occur near the surface even if the resonance condition 
$H_i=0$ is not satisfied in the bulk. The relaxation can propagate and 
possibly become total as $t\to\infty$.
Whether this occurs or not is one of the questions to be answered by the
simulations reported below.

\begin{figure}[hbt]
\begin{center}
\includegraphics[bbllx=35pt,bblly=50pt,bburx=530pt,bbury=460pt,
width=7 truecm ,angle=0]{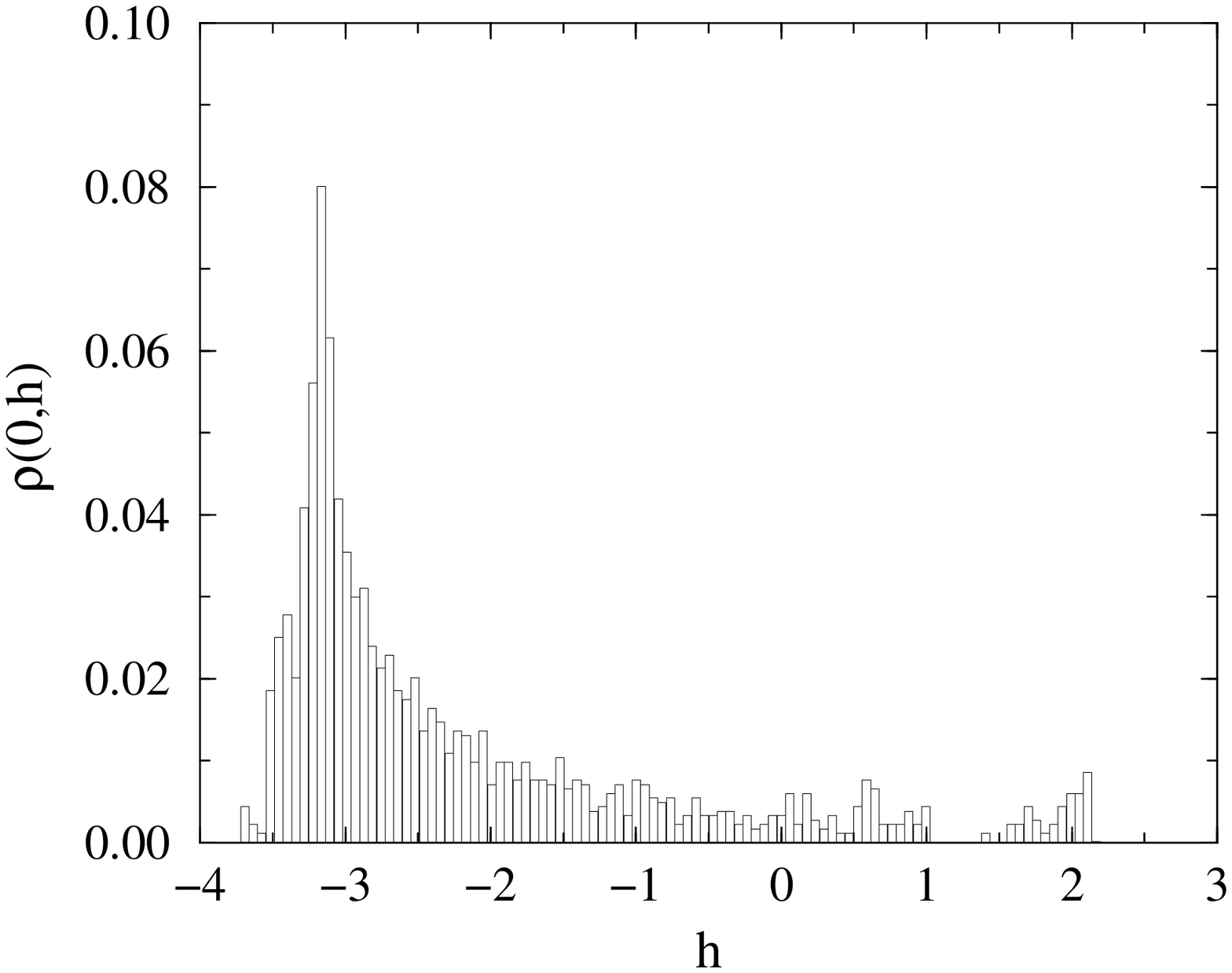}
\end{center}
\begin{center}
\includegraphics[bbllx=20pt,bblly=100pt,bburx=575pt,bbury=690pt,
width=7 truecm ,angle=270]{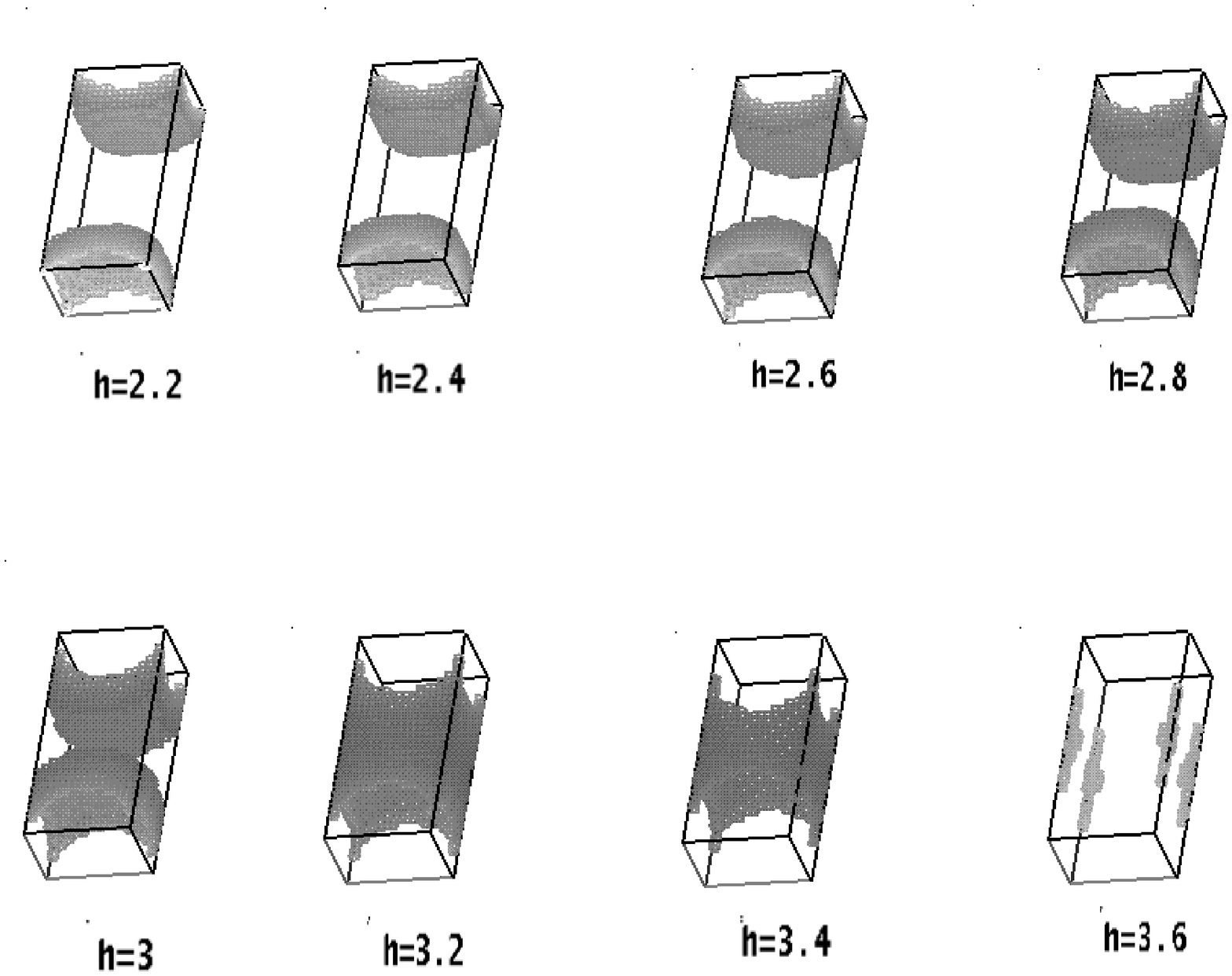}
\end{center}
\caption{a) Distribution of local fields
at the sites of a simple cubic
lattice occupied by spins parallel to the $z$ axis,
for a parallelepiped of $ 12 \times 17   \times 36 $ spins.
b) The local fields through this sample: below each picture is reported 
the value of the external field to be applied to bring the shaded 
spins to resonance (the local field is given in units 
$g\mu_BS/a^3\simeq 80$ Oersted, see text in Sect. \ref{ch4}).}
\label{t=00}
\end{figure}

If the sample is not a sphere (nor an ellipsoid) the resonance
condition can only be satisfied in a part of the sample, even at
$t=0$, when $S_i=-s$ (Fig. \ref{t=00}b) so that the magnetization has
a broad distribution (Figs.  \ref{t=0}c,\ref{t=00}a). 
\noindent According to Prokofiev and Stamp~\cite{prok}, the variation of the
magnetization at short time $t$ is still proportional to
$\sqrt{t/\tau}$ as in (\ref{3.5}); such prediction has recently
received beautiful experimental confirmations~\cite{ohm,wern}.
Moreover in Ref.~\cite{prok} is found that the short time relaxation
constant $\tau$ contains the volume where the local field has its
resonance value. This prediction ignores the possible extension of
this volume in time, and is therefore somewhat speculative, but it
opens the possibility to experimentally probe the field
distribution~\cite{wern}.

For a non-spherical sample, the square root law  is observed on 
a broader field interval since the field distribution is broader.
Moreover, the square root law
can be generalized to any value of the initial magnetization
namely~\cite{wern}
                                               
\begin{equation}
m(0)-m(t) \approx \hbox{Const} \times \sqrt{t}~.
\label{3.55}
\end{equation}
                                               
This result holds only if the density
of the local field does not change much 
for a field variation of the order of the 
resonance width $H_1$. This excludes spherical samples.

Some of the formulae written in section \ref{ch3} for a spherical
sample can be generalized. For instance, if one introduces the local
magnetization $m(\vec r, t)$ and the local distribution $P(\vec r, t;
H)$ of internal field, (\ref{3.1}) can be generalized as

\begin{equation}
\frac{\partial}{\partial t}m(\vec r,t)
=-{4\over 3}\eta(0) H_1  P(\vec r, t;0)m(\vec r,t)~.
\label{3.111}
\end{equation}
       
Integrating  $P(\vec r, t; H)$ on $\vec r$, one obtains 
the distribution of local fields at time $t$,

\begin{equation}
\rho(t;H) = \int d^3r \int_{-\infty}^\infty dH' P(\vec r, t;H) \delta(H'-H)
\label{3.111'}
\end{equation}
which is shown for $t=0$ for various sample shapes on Figures 
\ref{t=0}, \ref{t=00} and \ref{t=000}.

\section{Monte-Carlo simulations: model and method}
\label{ch4}

The model sketched in section \ref{ch2} is not yet completely defined. 
The crystal lattice, for instance, has not been defined.

The material Fe$_8$ has a complicated, triclinic crystal structure. It
can be taken into account in the simulations, but this does not
warrant realism. A complete, realistic calculation would imply a
calculation of the coefficients $g_{ij}$ taking into account the exact
shape of the electronic wave functions. This would be a difficult
task. Moreover, the presence of short range exchange interactions,
which would be more difficult yet to calculate, cannot be excluded.

\begin{figure}
\begin{center}
\includegraphics[bbllx=0pt,bblly=0pt,bburx=540pt,bbury=450pt,
width=7 truecm ,angle=0]{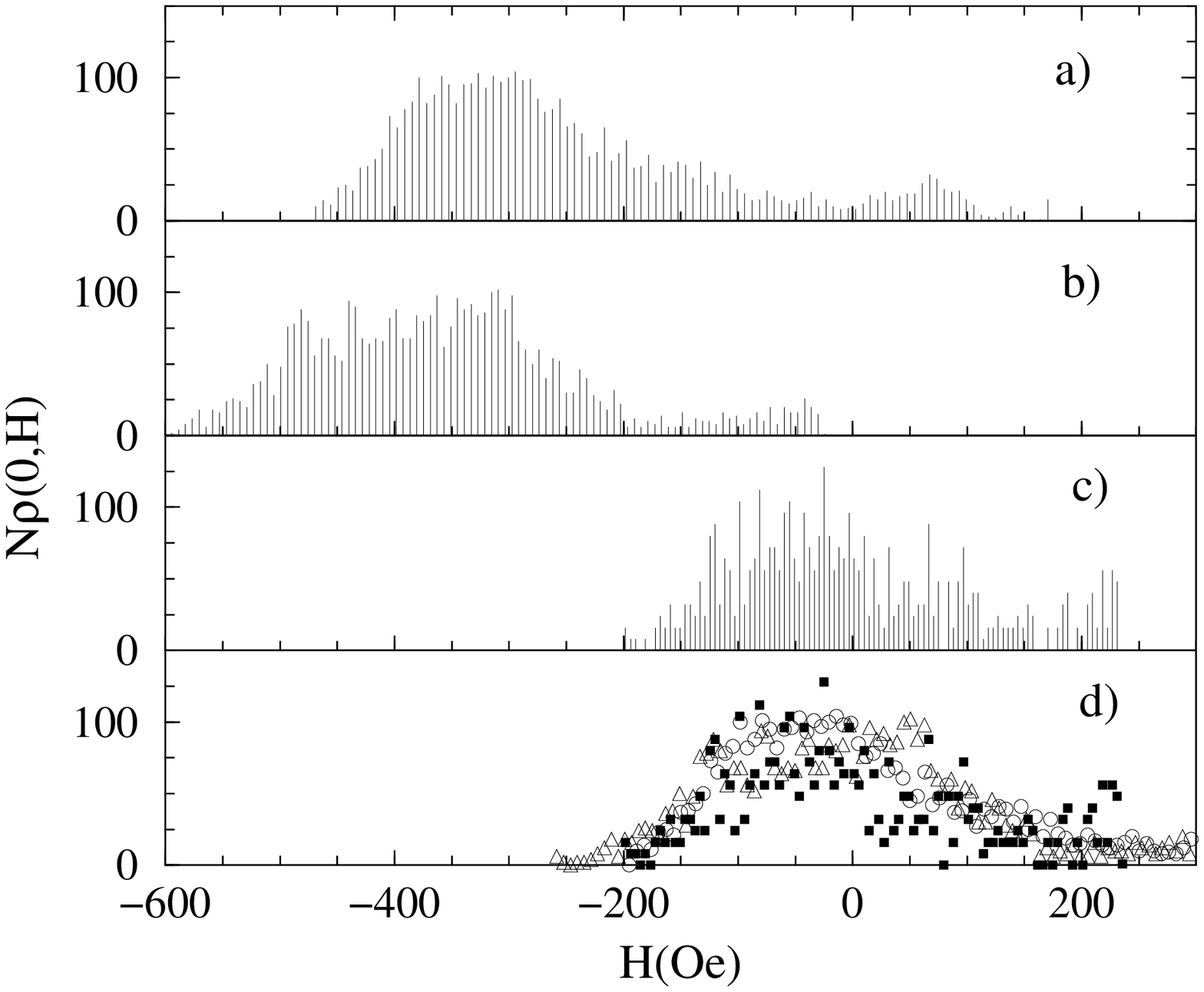}
\end{center}
\caption{Distribution of internal fields in a magnetically saturated 
sample of cubic shape.
a) Spins localized at the Fe-ions sites of the actual material, 
ignoring the actual electronic density.
b) Single giant spins localized at the sites of a triclinic lattice 
with the unit cell parameters of \protect{Fe$_8$}. 
c) Single giant spins localized at the sites of a cubic lattice with
lattice constant \protect{$a=13~\mbox{\AA}$}.
d) The same distributions given in the previous picture, with a) (open 
squares) and b) (open triangles) shifted by 280 and 360 Oersted, respectively.}
\label{t=000}
\end{figure}

We believe that all these unknown effects can be taken into account by
a single parameter, which is the value $H_0$ of the external field at
resonance for a sphere or at the center of a cube.  For a cubic
sample of a cubic crystal, $H_0=0$ by symmetry. 
For the real material Fe$_8$, $H_0$ does
not vanish, cannot be calculated for the reasons explained above, but
can easily be measured. When it is known, the field distribution in a
sample of any shape can be calculated for a cubic lattice, and then
obtained for the real material by shifting the field scale by an
amount $H_0$.
The above views are supported by figure \ref{t=000} which shows the 
internal field distribution for a cubic lattice compared to other possible
models. The distributions are not very different apart from a shift.

Most of our simulations have been performed on a
cubic lattice.  We have done a single simulation on a system with the
crystal structure of Fe$_8$ (but with localized spins) and checked
that there is no significant difference with simulations on a cubic
lattice

An important parameter is the resonance width $H_1$. If it were larger
than $H_{\rm dip}$, the relaxation would be fast and exponential. The
experimentally observed, slow relaxation is certainly related to a
small value of $H_1$. In the simulations, $H_1$ cannot be
smaller than the typical distance between the discrete
 values of the dipole
field, which has a lower limit since the sample cannot be very 
large (see Fig.~\ref{t=0}c); however, the sample dimension in our
simulation allows us to employ a value of $H_1$ of the same order of
magnitude of that deducible from experiments. In evaluating the local
field we add to the dipolar contribution given by Eqs. (\ref{2.1'}) and
(\ref{2.2}) a constant, external, applied field which can be
varied. In most of the figures reporting results of simulations 
we use the reduced field $h=H/\overline H$, where 
$\overline H=g\mu_BS/a^3\simeq 80$ Oersted, for $S=10$ and $a=13~\mbox{\AA}$.

The Monte Carlo algorithm employed to simulate the dynamics of our
systems is an implementation of those proposed and discussed 
elsewhere~\cite{AdamBL97,Binder79}.

We start from a completely magnetized sample at time $t=0$, and we
evaluate the local field $H_i$ acting on any site $i$ of the lattice;
we then cycle through the following steps:

{\it i)} We single out those sites where $|H_i|\leq H_1$; let us
  denote with $n_0$ the number of such spins which, according to our
  model, can relax with a probability given by Eq.(\ref{2.3}).
{\it ii)} We increment time replacing $t$ by $t+\Delta t$, where
  $\Delta t$ is chosen stochastically with probability 
$\eta_0 n_0 e^{-\eta_0 n_0\Delta t}$, i.e. we set $\Delta
t=-\ln\xi/(\eta_0 n_0)$, where $\xi$ is a generated number uniformly
distributed in $(0,1)$.
{\it iii)} We randomly choose one of those $n_0$ spins singled out in
  step $i)$ and flip it with probability given by Eq.\ref{2.3}.
{\it iii$_a$)} If the spin has been flipped we update the total
magnetization and the fields on all sites of the lattice.
{\it iv)} We come back to step {\it i)}.

For any set of the simulation parameters different independent runs
were made and averaged. Some sample runs where also done using the 
more elementary algorithm
which uses constant time steps; the results obtained are the same, but 
it takes much longer time, due to the smallness of the time step to be 
used to get stable results.

\section{Monte-Carlo simulations: results}
\label{ch5}

\subsection{Short time behavior}
\label{ch5-2}

\subsubsection{Spherical samples}
\label{ch5-2-14}

\begin{figure}
\begin{center}
\includegraphics[bbllx=25pt,bblly=25pt,bburx=540pt,bbury=440pt,
width=7 truecm ,angle=0]{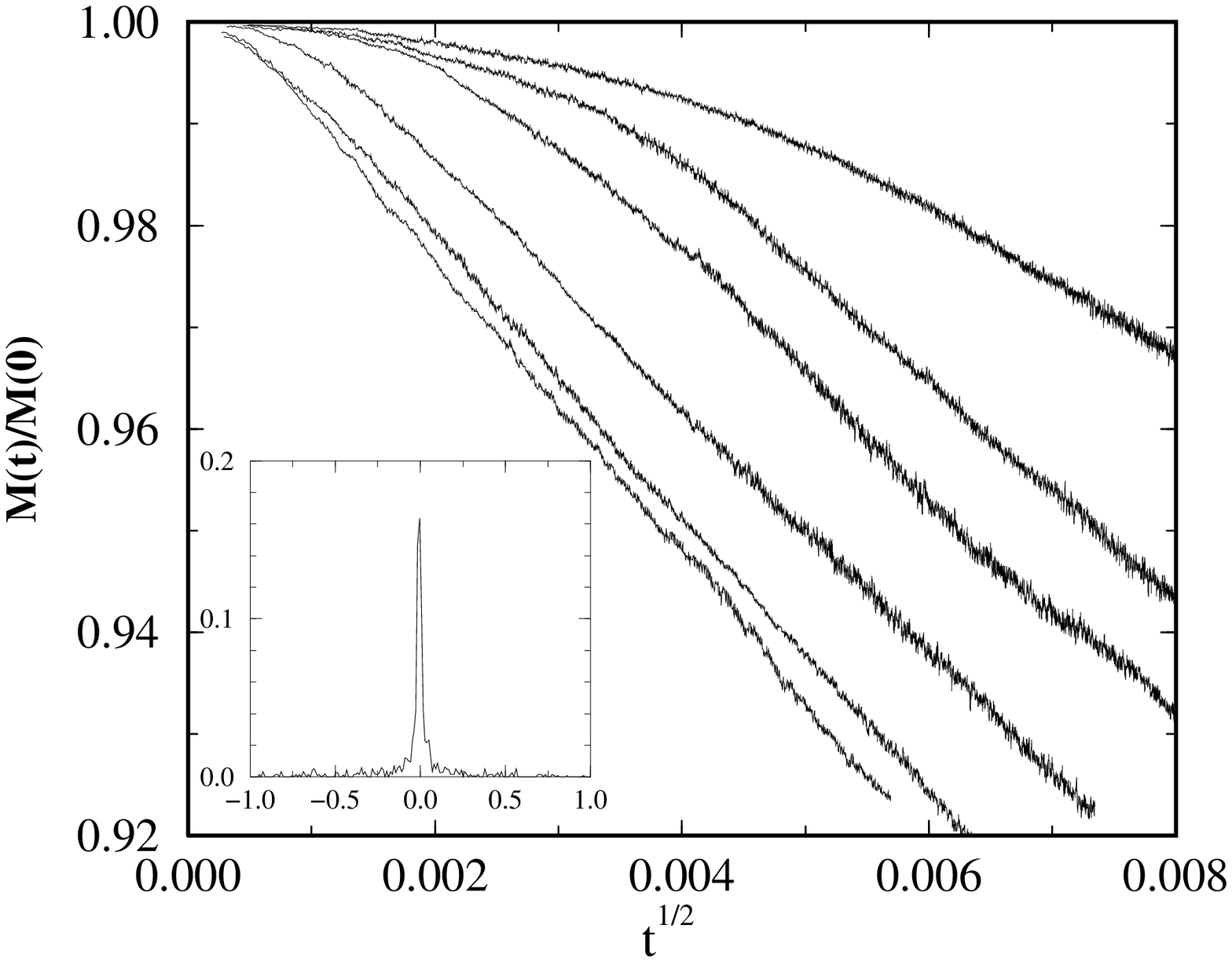}
\end{center}
\caption{Magnetization as a function of the square root of the
time for a spherical sample of radius 13 lattice constants for 
$\eta(0)=10^4$ and various applied fields h.
From the uppermost curve: h=0.3, h=0.2, h=0.15, h=0.1, h=0.05,
h=0. The reported data are the average over 10 independent runs.
In the inset the field distribution is reported.
}
\label{sfer}
\end{figure}

For a spherical sample, the Prokofiev-Stamp formula (\ref{3.5}) is
very well satisfied if the uniform, initial local field $H(0)$ is
smaller than the width $H_1$. 
(Fig. \ref{sfer}). For those fields, 
the agreement with (\ref{3.5}) is good (except at
very short times) as long as the magnetization is larger than 80\% of
its initial value.
At extremely short times, the
magnetization decays linearly with $t$ as well known~\cite{prok}. 
This can be understood because
the spins relax independently since the average distance between
reversed spins is very large and the dipole field created by the
reversed spins is smaller than the width $H_1$.

Away from the resonance, i.e. for $H(0)=H>H_1$, formula (\ref{3.5}) is
no longer satisfied.  However, if the initial field is not too far
from the resonance, the magnetization becomes a linear function of
$\sqrt{t}$ after a certain time $\tau_1(H)$. This phenomenon can
presumably be interpreted as follows. For short times, $P(\vec r,
t;0)=0$ except near the surface, so that $\partial m/\partial t$ is
small according to Eq. (\ref{3.111}).  However, a partly relaxed zone,
with $P(\vec r, t;0)\neq 0$, will progressively invade the whole
sample. The invasion is total at time $\tau_1(H)$.  This time can be
expected to become very long for big samples.

\subsubsection{Parallelepipedic samples}
\label{ch5-2-2}

\begin{figure}
\begin{center}
\includegraphics[bbllx=25pt,bblly=25pt,bburx=540pt,bbury=440pt,
width=7 truecm ,angle=0]{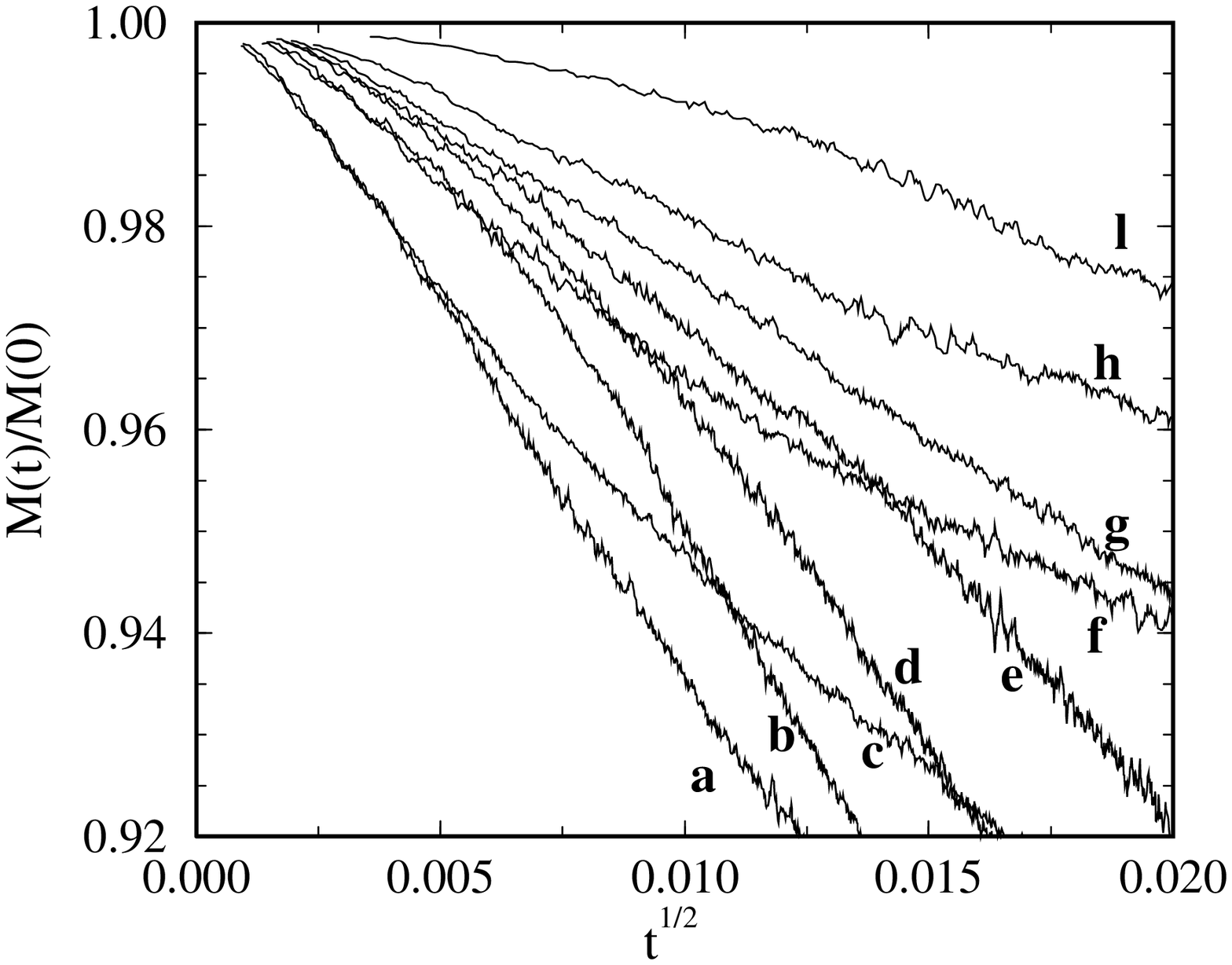}
\end{center}
\caption{Magnetization as a function of the square root of the
time for a parallelepipedic sample of $12 \times 17 \times 36$ spins
for $\eta(0)=10^4$; the longest side is along the easy magnetization axis. 
The various curves correspond to the following applied fields:
a) h=3; b) h=2.6; c) h=3.2; d) h=2.4; e) h=2.2; f) h=3.4; g) h=2; h)
h=1.6; l) h=3.6. The reported data are the average over 10 independent runs.
}
\label{parall}
\end{figure}

We investigated an elongated, parallelepipedic sample whose shape ($
12 \times 17 \times 36 $ spins) roughly corresponds to that which was
experimentally measured~\cite{Sangrego}.  The Prokofiev-Stamp
prediction (\ref{3.55}) is satisfied for short times as shown by
figure \ref{parall}.  However, these `short' times are differently
short for different external fields. For certain field values, the
slope $|dm/d\sqrt{t}|$ suddenly decreases after a rather short times
and the magnetization curve crosses other magnetization curves which
satisfy the Prokofiev-Stamp prediction on a longer interval.  This
crossing has not been experimentally seen.  It is of interest to
relate the different shapes of the demagnetization curve to the
initial internal field density $\rho(0;H)$ shown by Fig. \ref{t=00}a.
The curve $\rho(0;H)$ has a sharp maximum, decreases abruptly to 0 on
one side and much more smoothly on the other side. The field values
which satisfy the Prokofiev-Stamp prediction for a long time
correspond to the smooth side. Those which correspond to the steep
side and to the maximum satisfy the Prokofiev-Stamp prediction during
a shorter time.  We have no precise explanation for these
observations, but it seems to be related to the sharp edge of the
distribution and to the fact that the spin interested by relaxation
are in this case mostly concentrated on the surface of the sample.

\subsection{Long time behavior}
\label{ch5-1}

\subsubsection{Spherical sample}
\label{ch5-1-1}

\begin{figure}
\begin{center}
\includegraphics[bbllx=50pt,bblly=40pt,bburx=520pt,bbury=440pt,
width=7 truecm ,angle=0]{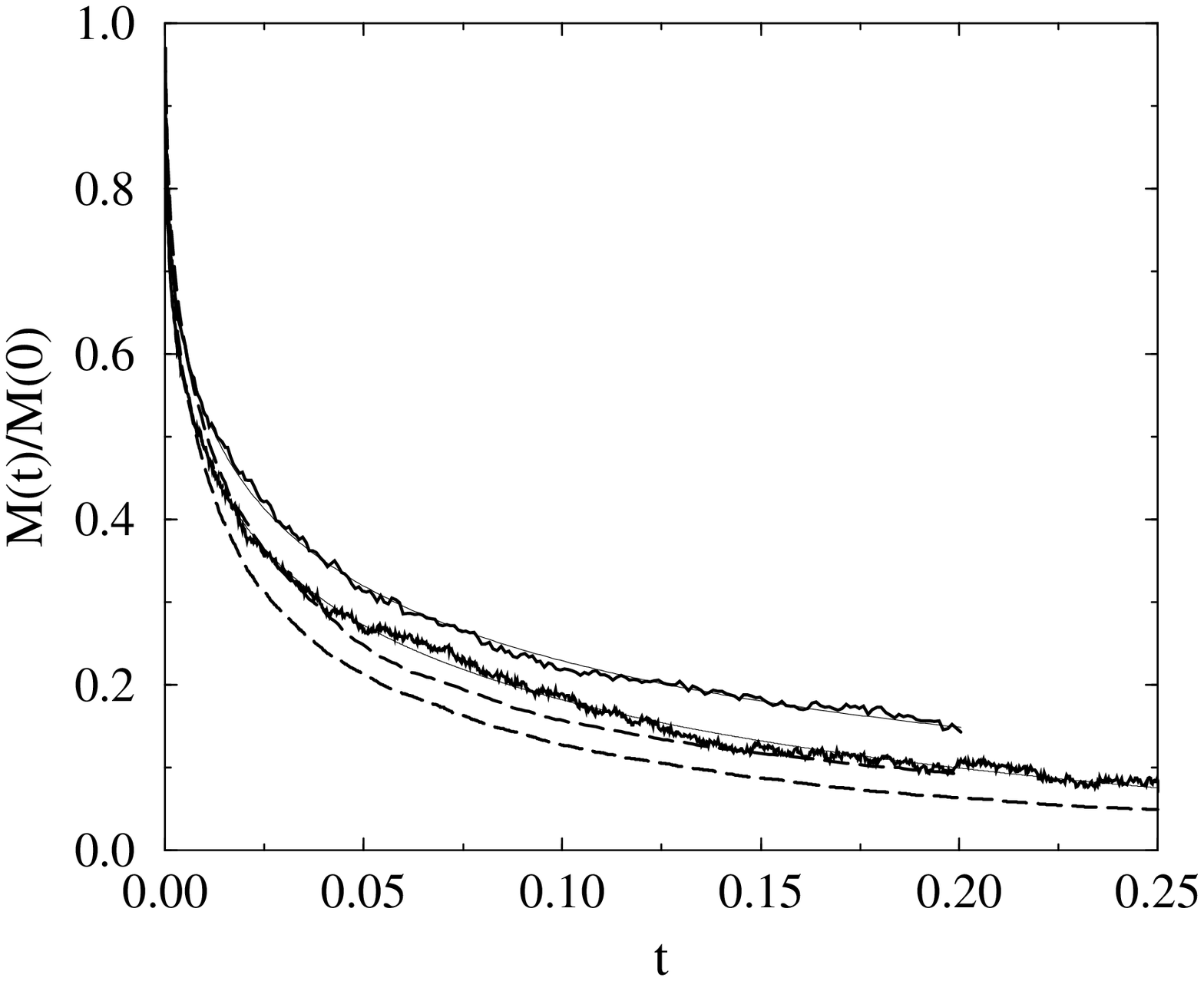}
\end{center}
\caption{Long time relaxation of the magnetization in a spherical
  sample of radius $r=13$ for $\eta(0)=10^4$. 
  Continuous lines: magnetization as function 
  of time for external applied field $h=0.1$ ( uppermost curve, fitted 
  stretched exponential parameters: $\beta_1=0.36$, $\tau=0.036$) and
  $h=0$ (fitted stretched exponential parameters: $\beta_1=0.33$, 
 $\tau=0.037$). The dashed lines show the fraction of spins which have 
 never flipped, for the same values of the applied field and in the 
 same order. The thin, steady lines through the magnetization curves are the
 stretched exponential fits.
}
 \label{sfer'}
\end{figure}

Figure \ref{sfer'} shows the total magnetization at long times in the
case of a sphere. It turns out to be well fitted by a stretched
exponential with $\beta_1=0.37$; however the good agreement with
(\ref{1.4}) is not very significant since, as will be seen later, the
parameter $\beta_1$ seems to depend on the sample shape, and the
experiments are mostly done on a parallelepipedic sample. No good fit
by (\ref{3.4'}) is possible.
                                
\begin{figure}
\begin{center}
\includegraphics[bbllx=15pt,bblly=15pt,bburx=575pt,bbury=500pt,
width=7 truecm ,angle=0]{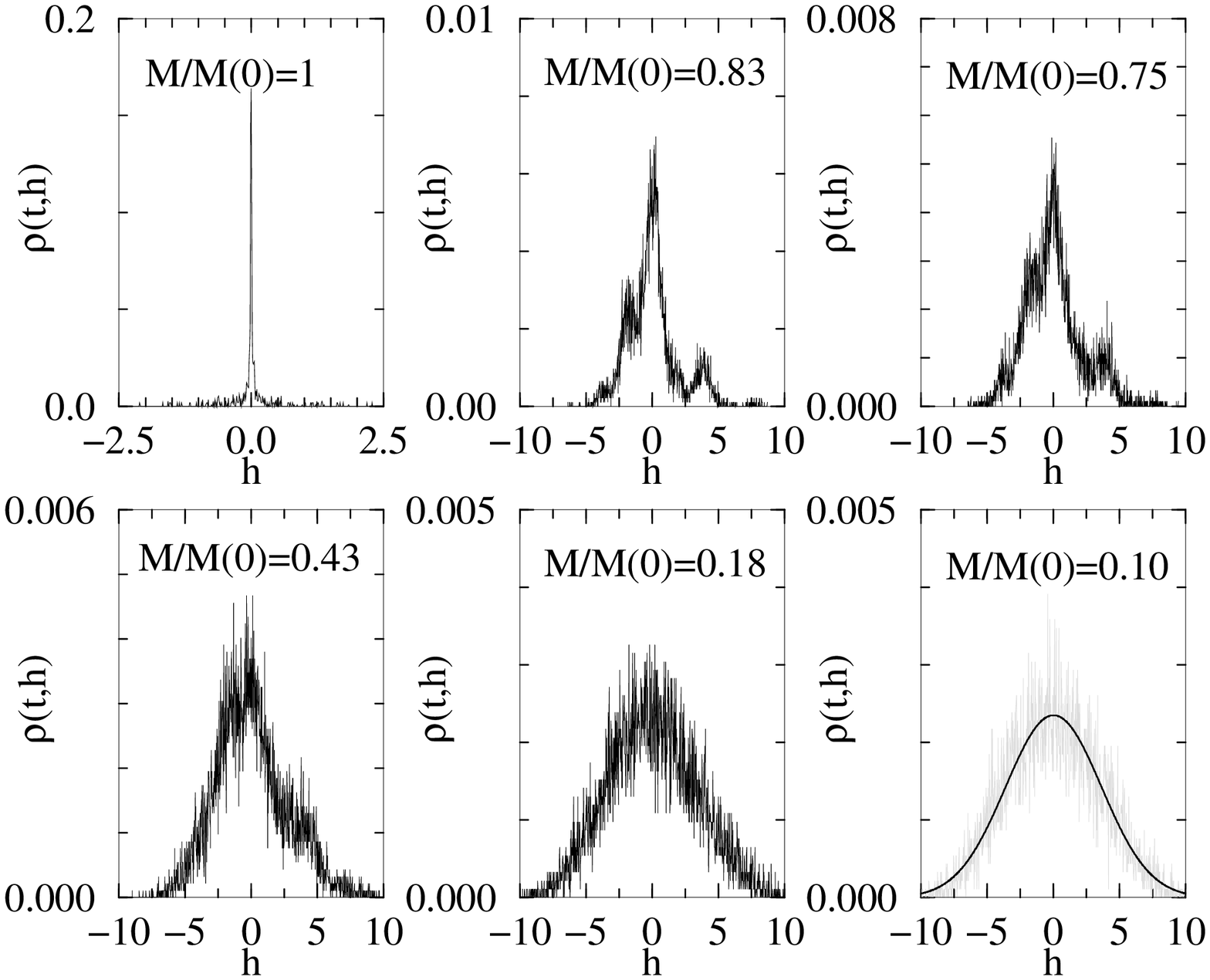}
\end{center}
\caption{Evolution of the field distribution in a sphere with no applied
  field; in each picture the value 
  of the corresponding magnetization is given.
}
 \label{fieldevolsph}
\end{figure}

The distribution $\rho(t;H) $ of the internal fields, initially very  
sharply peaked, smoothens as shown by Fig. \ref{fieldevolsph}.

\subsubsection{Cubic and parallelepipedic samples: effect of geometry.}
\label{ch5-1-2}

\begin{figure}
\begin{center}
\begin{center}
\includegraphics[width=7 truecm]{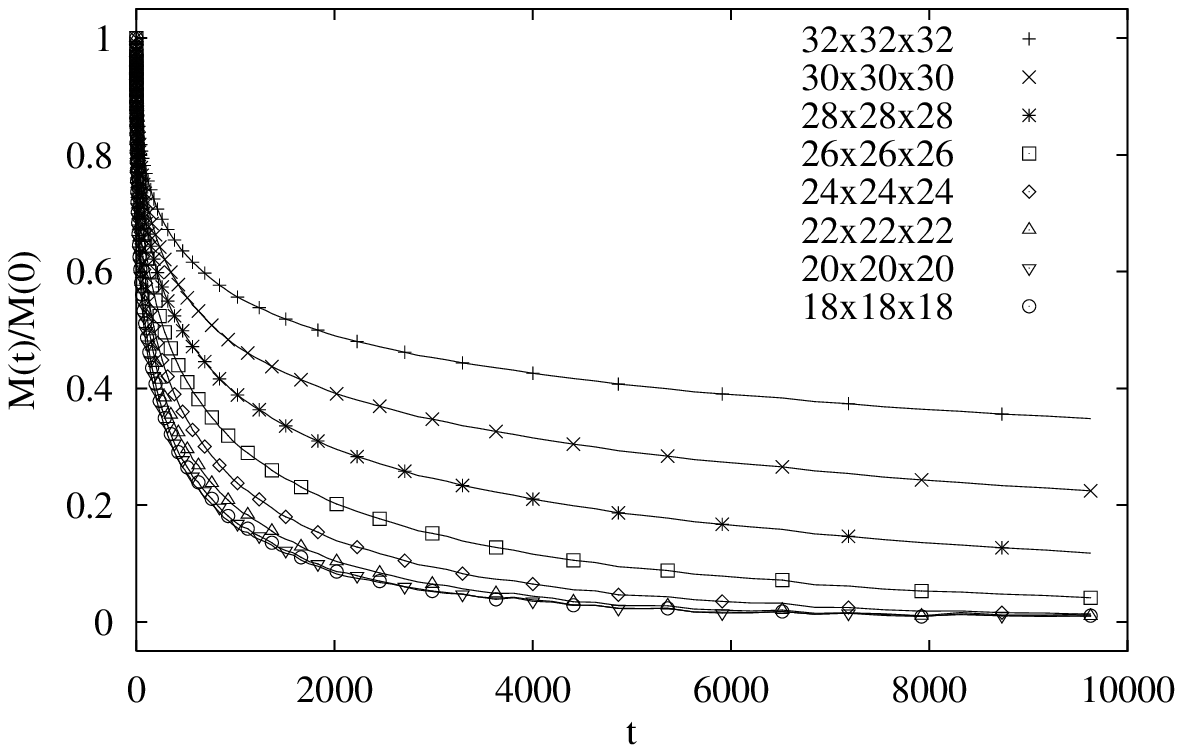}
\end{center}
\caption{Magnetization as a function of time for a cubic sample of 
$32^3$ spins
and in regions of cubic shape having the same center as the cube.
The external field is chosen such that the internal field vanishes
at the center of the sample at $t=0$, $H_1=0.1$, $\eta(0)=1$ (in order 
to compare with the data for other sample shapes reported in the other 
figures, please note that a change of $\eta(0)$ entails only a
rescaling of time),
and the average is done over 20 realizations. 
Upper curve: magnetization in the full sample of $32^3$ spins.
Lower curves: magnetization in sub-cubes, concentric to the full one, of 
the dimension given in the legend.
}
 \label{erwan}
\end{center}
\end{figure}

The magnetization of a cube is shown by Fig. \ref{erwan}, together with
the magnetization of smaller cubes of various sizes having the same center as the whole
sample. The external field $H$ is chosen such that the internal field vanishes
at the center of the sample at $t=0$, i.e. $H=0$ in our model. 
The magnetization of the central region is seen to have almost completely vanished
while the relaxation is still very weak at the periphery of the sample. This suggests that 
the latest stage of the relaxation is dominated by the motion of the boundary of
the relaxed region.

The cubic shape of the inner regions chosen in Fig. \ref{erwan} is the
simplest choice, but perhaps not the best. At short times, the relaxed
region is indeed expected to be the volume where the internal field
vanishes in the saturated sample, and this volume is more complicated
than a cube.

The relaxation of the central region is well described by the formula
\begin{equation}
t = A \ln^2 [m(t)/s] + B \ln^3 [m(t)/s]~,
\label{serie}
\end{equation}
with $B/A\approx 0.1$.

\begin{figure}
\begin{center}
\includegraphics[bbllx=50pt,bblly=40pt,bburx=520pt,bbury=440pt,
width=7 truecm ,angle=0]{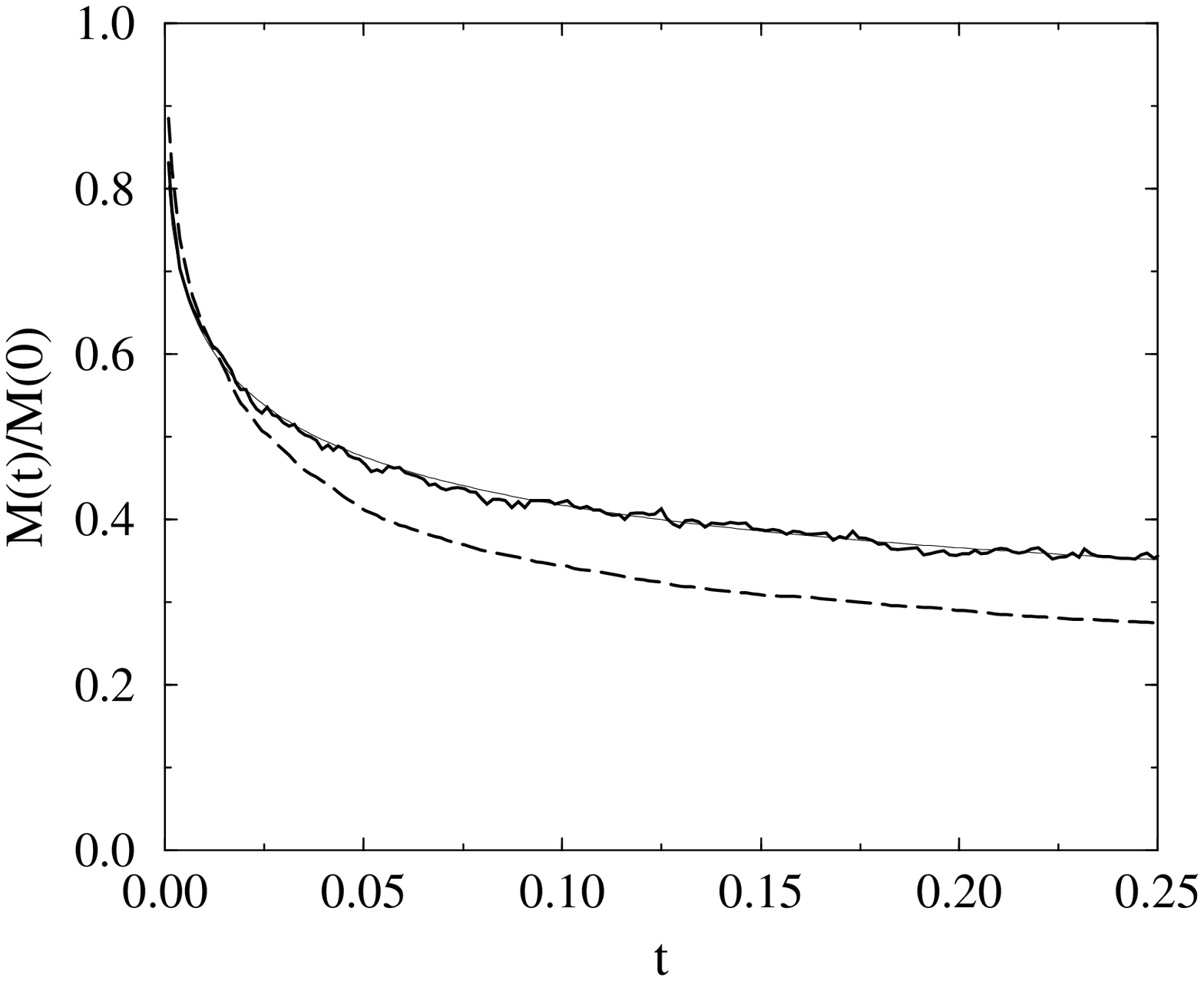}
\end{center}
\caption{Long time relaxation of the magnetization in a parallelepipedic
  sample $12\times17\times36$ for $\eta(0)=10^4$ in external applied 
  field $h=2.4$. 
  Continuous line: magnetization as function of time; dashed line: 
 fraction of spins which have 
 never flipped; the thin, steady line through the magnetization curves is the
 stretched exponential fit with parameters $\beta_1=0.33$, $\tau=0.026$.
}
 \label{parall'}
\end{figure}

The decay of the total magnetization is very well fitted, until 0.15 times
the saturation value, by a stretched exponential with $\beta_1=1/4$,
which is quite far from the experimental value (\ref{1.4}) for
elongated samples.  A simulation done in an elongated parallelepipedic
sample (Fig. \ref{parall'}) yields $\beta_1=0.33$, which is closer to
the experimental value.

\begin{figure}
\begin{center}
\includegraphics[bbllx=15pt,bblly=15pt,bburx=575pt,bbury=500pt,
width=7 truecm ,angle=0]{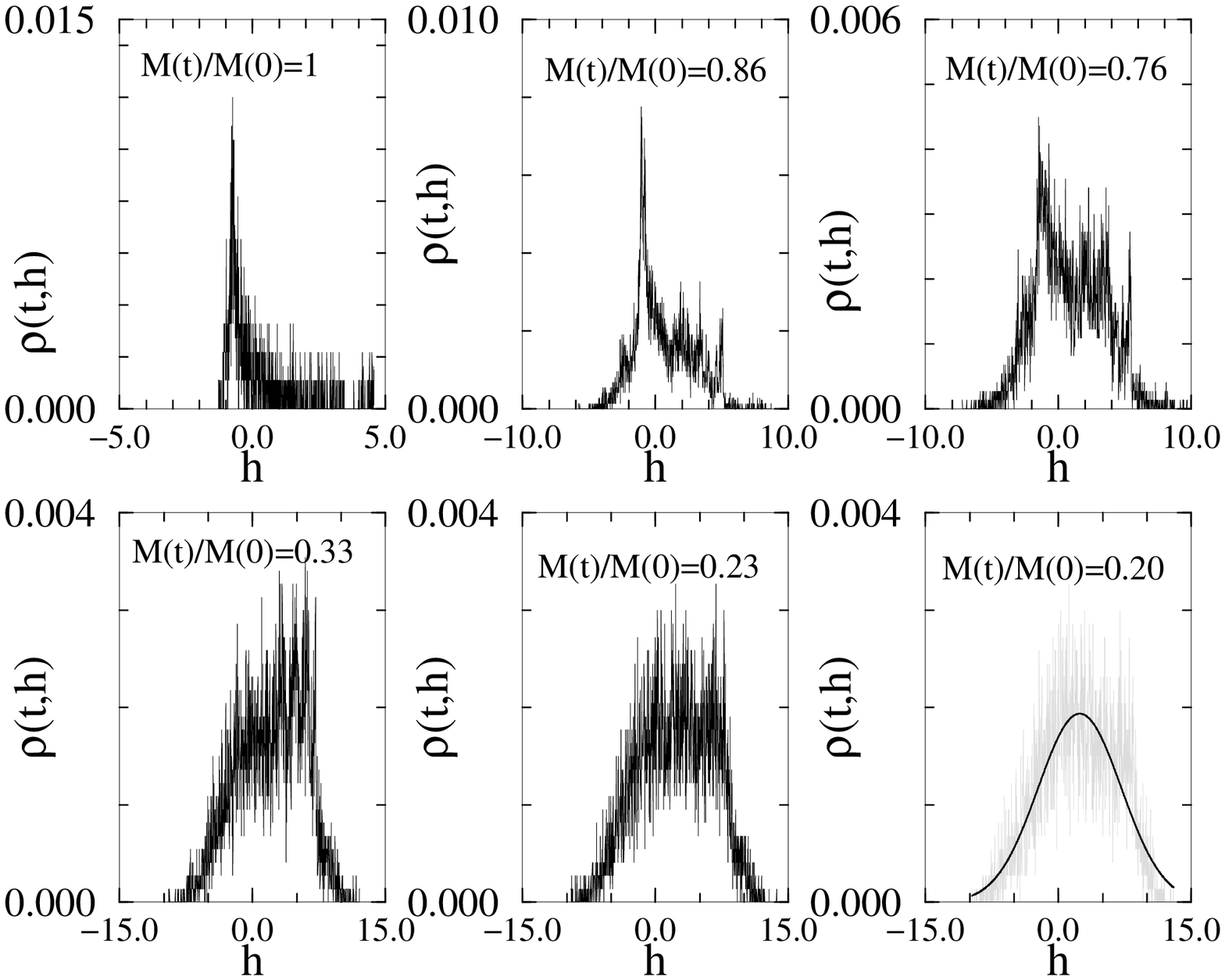}
\end{center}
\caption{Evolution of the field distribution in a parallelepiped 
  $12\times17\times36$ with
  external applied field $h=2.4$, corresponding to a local field
  $h_i=-0.704$ at the center of the sample; in each picture the value 
  of the corresponding magnetization is given.
}
 \label{fieldevolpar}
\end{figure}

The distribution $\rho(t;H) $ of the internal fields, initially very
sharply peaked, smoothens as shown by Fig. \ref{fieldevolpar} in the
case of a parallelepiped.

\subsubsection{Effect of $H_1$}
\label{ch5-1-3}

\begin{figure}
\begin{center}
\includegraphics[bbllx=25pt,bblly=25pt,bburx=560pt,bbury=490pt,
width=7 truecm ,angle=0]{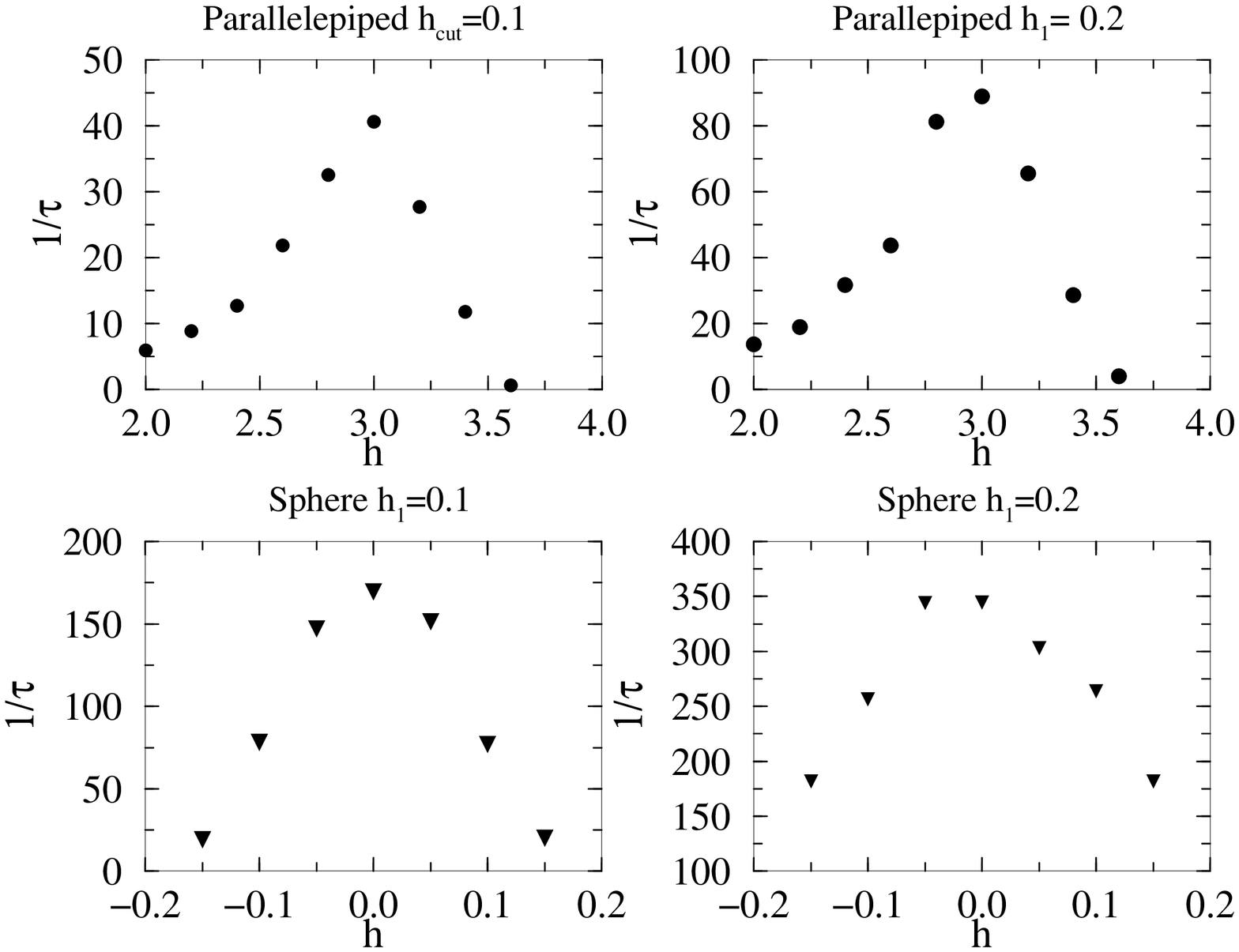}
\end{center}
\caption{Short time inverse relaxation time  
$\tau^{-1}$ as a function of applied
  field for a sphere of radius $r=13$ (lower pictures) and a
  parallelepiped $12\times17\times36$ for two different widths of the
  resonance, $H_1=0.1$ (left) and $H_1=0.2$ (right) and $\eta(0)=10^4$. 
}
 \label{H_1}
\end{figure}

The effect of $H_1$ is shown by Fig. \ref{H_1}. The  time scale is modified as expected, 
but the dependence of the relaxation time
with respect to the external field is not strongly modified, and well
compares with that observed in the experiments.

\section{Conclusions}
\label{ch6}

Our contribution complements the work of Prokofiev \&
Stamp~\cite{prok,prok'} and suggests new experiments.  In particular,
it is shown that the overall magnetization is remarkably well fitted
by a stretched exponential, with an exponent which seems to depend on
the sample shape and takes values from 1/4 to 0.4. Moreover, it is
pointed out that the slow relaxation observed in parallelepipedic
samples is mainly a result of the slow extension of the relaxed
region, and the central region relaxes much more rapidly. The
experimental check of this property would be a crucial test of the
model. It might be feasible using multisquids as in the recent
experiment of Wernsdorfer et al.~\cite{wern}.

\begin{acknowledgement}

Fruitful discussions with D. Gatteschi and R. Sessoli are gratefully
acknowledged. 

\end{acknowledgement}


\begin{thebibliography}{}

\bibitem{prok}  N.V. Prokofiev, P.C.E. Stamp, Phys. Rev. Lett. {\bf 80}, 5794 (1998). 

\bibitem{prok'} N.V. Prokofiev, P.C.E. Stamp, J. Low Temp. Phys {\bf 113}, 1147 (1998).

\bibitem{Sangrego} C.Sangregorio, T. Ohm, C. Paulsen,
R. Sessoli, D. Gatteschi, Phys. Rev. Lett. {\bf 78 },  4645 (1997).

\bibitem{wern} Wernsdorfer, W., Ohm, T., Sessoli, R., Paulsen, C.
to be published  (1999).

\bibitem{ohm} T. Ohm, C.Sangregorio, C. Paulsen, Eur.  Phys. J. B 
{\bf 6}, 195 (1998).

\bibitem{HB} F.Hartmann-Boutron, P.Politi and J.Villain, Int. Journal of
 Modern Phys. B {\bf 10}, (21), 2577-2637 (1996).

\bibitem{OhmPhD} T. Ohm, PhD Thesis, Univ. of Grenoble (1998).

\bibitem{AdamBL97} E. Adam, L. Billard and F. Lan{\c c}on, Phys. Rev. E
  {\bf 59}, 1212 (1999).    

\bibitem{Binder79} K. Binder in {\it Monte Carlo Methods in
    Statistical Physics}, Ed. K. Binder, Springer Verlag (1979).


\end{thebibliography}
\end{document}